# Has Anti-corruption Efforts lowered Enterprises' Innovation Efficiency? — An Empirical Analysis from China

Liu lunwu and Liu shi

（School of Economics, Jiangxi University of Finance and Economics, Nanchang Jiangxi Province PRC）

【Abstract】This study adopts the fixed effects panel model and provincial panel data on anticorruption and the innovation efficiency of high-level technology and new technology enterprises in China from 2005 to 2014, to estimate the effects of the anticorruption movement on the innovation efficiency of enterprises at different corruption levels. The empirical results show that anticorruption is positively correlated with the innovation efficiency of enterprises; however, the correlation is differentiated according to different corruption levels and business natures. At a high level of corruption, anticorruption has positive impacts on enterprises' innovation; at a low level of corruption, it negatively affects innovation efficiency. However, anticorruption has negative effects on the innovation efficiency of state-owned enterprises at both high and low corruption levels; for nonstate-owned enterprises, its effects are positive at a high corruption level and negative at a low corruption level. The effects remain the same across different regions.

## 1.Introduction

Since 2010, the growth rate of China's economy has continuously decreased to less than double the number of digits for several quarters, during which time the country has entered the structural deceleration phase (Lin & Wang, 2016)[1], a critical period for transformation and upgrading. Increasing investment in R&D, improving innovation efficiency and realizing supply innovation are the keys to realizing the transformation of economic structure and sustainable and stable growth in China. However, because of the long history of " reciprocal rent-seeking " between governments and businesses in China, many enterprises can easily make profits by establishing special political connections with the government; as a result, innovation incentives are weakened. In addition, R&D investment is very limited and difficult to fully and reasonably use, so the overall innovation ability of enterprises is weak, and innovation efficiency is low in China (Li Dang et al, 2015).[2] Rent-seeking corruption has become an important factor affecting the innovation efficiency of Chinese enterprises and the most important obstacle to the sustainable growth of China's economy (Wang & Wang, 2016).[3] After the 18th National Congress of the Communist Party of China, strong anticorruption action strictly controlled the ties between government and businesses, and the cost of "reciprocal" rent-seeking between government and businesses increased sharply, which made it difficult for enterprises to obtain profits easily through political connections. According to the literature, some scholars believe that after anticorruption, some entrepreneurs (especially those of state-owned enterprises) experience "inertia" to protect themselves, to reduce their innovation and risk-taking activities. Some scholars believe that after this movement, some entrepreneurs (especially those from state-owned enterprises) exercise “political lethargy” and reduce their innovative, adventurous undertakings to keep themselves out of trouble; meanwhile, others believe that to survive, enterprises will increase R&D investment, improve innovation efficiency, and seek innovation to develop and grow for the sake of survival in the market. The exact impacts and intensity of impacts of the anticorruption campaign on the innovation efficiency of Chinese enterprises have yet to be scrutinized by this study.

According to previous studies, enterprises with political connections and corruption exist in most countries[4], and corruption lowers enterprises’ likelihood of engaging in product R&D[5]. This exerts greater impacts on patent acquisition[6] and negatively affects innovation efficiency[7] (Faccio, 2006; Felipe.Starosta. Waldemar, 2012; Fan & Xu, 2014; Caroline Paunov, 2016). On the other hand, some also believe that low-level corruption is the catalyst for business innovation and that high-level corruption is the stumbling block[8] (Li & Zhang, 2015). Under strict scrutiny and regulation, the number of enterprises with political connections decreases, and the corruption level decreases; thus, anticorruption increases the relative costs of these enterprises as well as innovation incentives (Faccio, 2006; Dang, Yang, & Yang, 2015). Most studies have argued that enterprises’ innovation is harmed by corruption and benefited by anticorruption and that the effects of corruption and anticorruption on innovation vary with enterprises of different sizes and natures. While many studies have focused on analyzing the relationship between corruption or anticorruption and enterprises' innovation, few studies have examined the effects of corruption or anticorruption on innovation efficiency, not to mention such effects at different corruption levels. Moreover, research on the relationship between anticorruption and enterprise innovation has focused mainly on analyzing the relationship between anticorruption and the R&D investment of enterprises. However, it is known that the amount of R&D investment does not fully represent the degree of innovation efficiency. Hence, the results of most of the empirical analyses cannot reflect the actual impacts of the anticorruption movement on the innovation efficiency of enterprises, which, as a matter of fact, are also influenced by different corruption levels. It is certain that the impacts differ across regions at different corruption levels, but how they differ has yet to be investigated.

Undoubtedly, it is important to quantify the effects of the anticorruption movement on enterprises' innovation efficiency using enterprise innovation efficiency as the dependent variable and anticorruption as the explanatory variable while considering different corruption levels and classifying enterprises based on their region and nature. It is believed that the following discussion can help crystallize the effects of anticorruption on the innovation efficiency and intensity of enterprises.

The following sections are structured as follows: the second section concerns modeling and data explanation; the third section describes the empirical analysis; the fourth section includes further discussion and robustness testing; and the last section provides the conclusion.

## 2. Modeling and Data Explanation

### (1) Modeling

First, the effects of the anticorruption movement on enterprises' innovation efficiency at different corruption levels are analyzed theoretically to propose analytic propositions and a theoretical framework.

Assuming II represents the net profit of enterprises, the function of II can be expressed as $\Pi = \mathrm{pQ} - C_P - \mathrm{C}_b - \mathrm{P}_r C_f - C_c$, where Q denotes the output level; P denotes the price level; $C_P$ is the operational cost; $\mathrm{C}_b$ refers to the corruption cost (represented by the number of bribes) and the function of the corruption level Cor and anti-corruption intensity Acor, represented as $\mathrm{C}_b = \mathrm{C}_b(\mathrm{Cor}, \mathrm{Acor})$; $C_f$ denotes the corruption penalty; $\mathrm{P}_r$ is the probability of discovering corruptive acts and the function of corruption level Cor and anti-corruption intensity Acor, represented as $\mathrm{P}_r = \mathrm{P}_r(\mathrm{Acor}, \mathrm{Cor})$; and $C_{Inv}$ is the allocated cost of innovation, also represented as the function of innovation efficiency Inv: $C_{inv} = C_{inv}(Inv)$. As the enterprises pursue the maximization of net profit, the solution is as follows:

$$\max \Pi = (\ \mathrm{p} * \mathrm{Q} - C_P - \mathrm{C}_b - \mathrm{P}_r * C_f - C_{Inv})$$

$$\mathrm{St}: C_{inv} > \mathrm{C}_b + \mathrm{P}_r * C_f \quad \text{Opt for rent-seeking}$$

$$\text{or } C_{inv} < \mathrm{C}_b + \mathrm{P}_r * C_f \quad \text{Opt for innovation}$$

The Lagrange function is constructed, and the first-order necessary conditions of variables Inv, Cor, and Acor are as follows:

$$\frac{\partial \pi}{\partial Inv} = -\frac{\partial \pi}{\partial C_{Inv}} * \frac{\partial C_{Inv}}{\partial Inv} + \lambda \frac{\partial \pi}{\partial C_{Inv}} * \frac{\partial C_{Inv}}{\partial Inv} = (1-\lambda) \frac{\partial \pi}{\partial C_{Inv}} * \frac{\partial C_{Inv}}{\partial Inv} \quad (1)$$

$$\frac{\partial \pi}{\partial Acor} = -\frac{\partial \mathrm{C}_b}{\partial Acor} - \frac{\partial \mathrm{P}_r}{\partial Acor} * C_f + \lambda \left( \frac{\partial \mathrm{C}_b}{\partial Acor} + \frac{\partial \mathrm{P}_r}{\partial Acor} * C_f \right) \quad (2)$$

We divide formula (2) by formula (1):

$$\frac{\partial \mathrm{Inv}}{\partial \mathrm{Acor}} = \frac{\left( \frac{\partial \mathrm{C}_b}{\partial Acor} + \frac{\partial \mathrm{P}_r}{\partial Acor} * C_f \right)}{\frac{\partial \pi}{\partial C_{Inv}} * \frac{\partial C_{Inv}}{\partial Inv}} \quad (3)$$

The left side of formula (3) represents how changes in anticorruption intensity affect enterprises' innovation efficiency. As the net profit and allocated cost of innovation are negatively correlated (i.e, $\frac{\partial \pi}{\partial C_{inv}} < 0$ ) and the allocated cost of innovation and innovation efficiency are positively correlated (i.e, $\frac{\partial C_{Inv}}{\partial Inv} > 0$), the denominator on the right side of formula (3) is $\frac{\partial \pi}{\partial C_{Inv}} * \frac{\partial C_{Inv}}{\partial Inv} < 0$. In the numerator of the fraction, $\frac{\partial \mathrm{C}_b}{\partial Acor} < 0$ and $\frac{\partial \mathrm{P}_r}{\partial Acor} * C_f > 0$. At a high level of corruption, $\left|\frac{\partial \mathrm{C}_b}{\partial Acor}\right| > \left|\frac{\partial \mathrm{P}_r}{\partial Acor} * C_f\right|$; in other words, after increasing the anticorruption intensity by one unit, the decrease in bribery is greater than the increase in penalty: $\left(\frac{\partial \mathrm{C}_b}{\partial Acor} + \frac{\partial \mathrm{P}_r}{\partial Acor} * C_f\right) > 0$, $\frac{\partial \mathrm{Inv}}{\partial \mathrm{Acor}} > 0$, and anticorruption and innovation efficiency are positively correlated. At a low level of corruption, $\left|\frac{\partial \mathrm{C}_b}{\partial Acor}\right| < \left|\frac{\partial \mathrm{P}_r}{\partial Acor} * C_f\right|$; in other words, after increasing the anticorruption intensity by one unit, the decrease in bribery is smaller than the increase in penalty: $\frac{\partial \mathrm{Inv}}{\partial \mathrm{Acor}} < 0$, and anticorruption and innovation efficiency are negatively correlated.

Based on the above analysis, two basic propositions can be proposed:

Proposition 1: When corruption is severe (at a high level), anticorruption positively affects enterprises' innovation efficiency, which means that anticorruption activities improve innovation efficiency.

Proposition 2: When corruption is mild (at a low level), anticorruption negatively affects enterprises' innovation efficiency, which means that the anticorruption movement harms the latter.

To further quantify the effects of anticorruption on innovation efficiency, the target variable—innovation efficiency—is expressed as $\mathrm{Inv}_{it}$, which is a function of its determinants, including the size of the enterprise, export intensity, and profit margin. Using these factors as the control variables ($Controls$), $\mathrm{Inv}_{it}$ can be expressed as follows:

$$Inv_{it} = \alpha_0 + Controls + \gamma_i + \delta_t + v_{it} \quad (4)$$

In this formula, $\alpha_0$ is the intercept, $\gamma_i$ is the entity fixed effect, $\delta_t$ is the time fixed effect, and $v_{it}$ is the random disturbance. To analyze whether anticorruption (movement) lowers enterprises' innovation efficiency, the anticorruption variable Acor is introduced into formula (4):

$$Inv_{it} = \alpha_0 + \alpha_{it} Acor_{it} + Controls + \gamma_i + \delta_t + v_{it} \quad (5)$$

In this formula, $Acor_{it}$ represents the anticorruption intensity. Formula (5) can be used to determine whether innovation efficiency is enhanced by a certain variable. It is decreased if $\alpha_{it} < 0$ and enhanced if $\alpha_{it} > 0$.

**(2) Data explanation**

The raw data in this study were obtained from the *China Statistical Yearbook*, the *Chinese Procuratorial Yearbook*, and the *China Statistical Yearbook on High and New Technology Industry (2005-2015)* between 2005 and 2015. The detailed explanation of the variable data is as follows:

The explained variable of the model is the innovation efficiency of enterprises. The corresponding statistics originate from raw data in the *China Statistical Yearbook on High and New Technology Industry (2005-2015)* regarding innovation investment capital $K$ (R&D expenses), innovation labor $L$ (the sum of workload calculated by the actual working time of full-time or part-time R&D personnel), and the innovation output $Y$ (sales revenue of new products) of high and new technology industries (consisting mainly of the manufacturing industries of aircraft, spacecraft, and related equipment, electronic and telecommunications equipment, computers and office equipment, medicine, and medical equipment and instruments). Stochastic frontier analysis and the translog production function are applied for estimation. The estimated results are shown in Table 1.

Table 1 Innovation Efficiency of High and New Technology Enterprises in 27 Chinese Provinces in 2005-2014

| Year | 2005 | 2006 | 2007 | 2008 | 2009 | 2010 | 2011 | 2012 | 2013 | 2014 | Mean |
|---|---|---|---|---|---|---|---|---|---|---|---|
| Beijing | 0.744 | 0.755 | 0.766 | 0.777 | 0.787 | 0.797 | 0.807 | 0.816 | 0.825 | 0.833 | 0.791 |
| Tianjin | 0.919 | 0.924 | 0.927 | 0.931 | 0.935 | 0.938 | 0.941 | 0.944 | 0.947 | 0.949 | 0.936 |
| Hebei | 0.094 | 0.106 | 0.119 | 0.134 | 0.149 | 0.166 | 0.183 | 0.199 | 0.218 | 0.237 | 0.161 |
| Shanxi | 0.211 | 0.229 | 0.248 | 0.267 | 0.286 | 0.306 | 0.326 | 0.346 | 0.367 | 0.387 | 0.297 |
| Inner Mongolia | 0.034 | 0.042 | 0.049 | 0.057 | 0.067 | 0.078 | 0.089 | 0.101 | 0.114 | 0.129 | 0.076 |
| Liaoning | 0.214 | 0.233 | 0.252 | 0.271 | 0.291 | 0.311 | 0.331 | 0.351 | 0.371 | 0.391 | 0.302 |
| Jilin | 0.144 | 0.159 | 0.176 | 0.193 | 0.211 | 0.229 | 0.248 | 0.267 | 0.287 | 0.307 | 0.222 |
| Heilongjiang | 0.048 | 0.056 | 0.066 | 0.076 | 0.087 | 0.099 | 0.113 | 0.127 | 0.142 | 0.157 | 0.097 |
| Shanghai | 0.661 | 0.676 | 0.689 | 0.703 | 0.716 | 0.729 | 0.741 | 0.753 | 0.764 | 0.775 | 0.721 |
| Jiangsu | 0.599 | 0.615 | 0.631 | 0.646 | 0.661 | 0.676 | 0.689 | 0.703 | 0.715 | 0.729 | 0.666 |
| Zhejiang | 0.267 | 0.287 | 0.306 | 0.326 | 0.347 | 0.367 | 0.387 | 0.407 | 0.427 | 0.447 | 0.357 |
| Anhui | 0.179 | 0.197 | 0.215 | 0.233 | 0.252 | 0.271 | 0.291 | 0.311 | 0.331 | 0.351 | 0.263 |
| Fujian | 0.670 | 0.684 | 0.698 | 0.711 | 0.724 | 0.736 | 0.748 | 0.759 | 0.771 | 0.781 | 0.728 |
| Jiangxi | 0.118 | 0.132 | 0.148 | 0.163 | 0.179 | 0.197 | 0.215 | 0.233 | 0.252 | 0.272 | 0.191 |
| Shandong | 0.471 | 0.489 | 0.508 | 0.527 | 0.545 | 0.563 | 0.581 | 0.597 | 0.614 | 0.629 | 0.552 |
| Henan | 0.193 | 0.211 | 0.229 | 0.248 | 0.268 | 0.287 | 0.307 | 0.327 | 0.347 | 0.368 | 0.279 |
| Hubei | 0.111 | 0.124 | 0.140 | 0.155 | 0.172 | 0.189 | 0.206 | 0.225 | 0.243 | 0.263 | 0.183 |
| Hunan | 0.191 | 0.208 | 0.226 | 0.245 | 0.264 | 0.284 | 0.304 | 0.324 | 0.344 | 0.364 | 0.275 |
| Guangdong | 0.479 | 0.498 | 0.516 | 0.535 | 0.553 | 0.571 | 0.588 | 0.605 | 0.621 | 0.637 | 0.560 |
| Guangxi | 0.087 | 0.099 | 0.113 | 0.127 | 0.141 | 0.157 | 0.174 | 0.191 | 0.208 | 0.227 | 0.152 |
| Hainan | 0.008 | 0.011 | 0.013 | 0.017 | 0.021 | 0.261 | 0.317 | 0.382 | 0.456 | 0.538 | 0.202 |
| Chongqing | 0.293 | 0.312 | 0.332 | 0.352 | 0.372 | 0.393 | 0.413 | 0.433 | 0.453 | 0.472 | 0.383 |
| Sichuan | 0.293 | 0.312 | 0.332 | 0.352 | 0.373 | 0.393 | 0.413 | 0.433 | 0.453 | 0.472 | 0.383 |
| Guizhou | 0.081 | 0.093 | 0.106 | 0.119 | 0.134 | 0.149 | 0.165 | 0.182 | 0.199 | 0.217 | 0.145 |
| Yunnan | 0.134 | 0.145 | 0.166 | 0.182 | 0.199 | 0.218 | 0.236 | 0.255 | 0.275 | 0.294 | 0.210 |
| Shaanxi | 0.098 | 0.111 | 0.125 | 0.139 | 0.155 | 0.172 | 0.189 | 0.206 | 0.225 | 0.243 | 0.166 |
| Gansu | 0.083 | 0.094 | 0.107 | 0.121 | 0.135 | 0.151 | 0.167 | 0.184 | 0.201 | 0.219 | 0.146 |
| Mean | 0.271 | 0.285 | 0.299 | 0.314 | 0.328 | 0.352 | 0.370 | 0.387 | 0.405 | 0.424 | |

Note: the above data were calculated and organized by the author.

The corruption level is represented by the number of registered offenders of bribery and corruption per 1000 government officials in each province, city, and autonomous region. The number of registered offenders is obtained from the *Chinese Procuratorial Yearbook*, whereas the number of officials refers to the number of employees in public administration units and social organizations in cities and towns registered in the *China Statistical Yearbook*. The greater the number of registered offenders per 1000 government officials is, the greater the corruption level. Moreover, the mean corruption levels of different provinces, cities, or autonomous regions in different years are computed. Values above the mean value are defined as manifestations of high corruption levels, and those below the mean represent low-level corruption.

Anticorruption intensity is expressed as the logarithm of the number of abuse-of-power criminal offenders under investigation. The greater the value is, the more vigorous the anti-corruption effort. The mean values of the anticorruption intensities of different Chinese provinces, cities, and autonomous regions in different years are also calculated.

The control variables include the enterprises' size, measured by the natural logarithm of the total firm profit; export intensity, measured by the ratio of aggregated export value to total firm profit; and profit margin, measured by the ratio of total profit to total revenue of main businesses. The data of the control variables are extracted from the *China Statistical Yearbook on High and New Technology Industry (2005-2014)*. The basic statistical description of the variables is illustrated in Table 2.

Table 2 Variable definitions and statistical descriptions

| Variable | Definition | Sample No. | Mean | SD | Minimum | Maximum |
|---|---|---|---|---|---|---|
| Inv | Innovation efficiency of enterprises | 270 | 0.350 | 0.242 | 0.008 | 0.949 |
| Cor | Corruption level: 1 = higher than the mean value and 0 = lower than the mean value | 270 | 0.456 | 0.498 | 0 | 1 |
| Acor | Anti-corruption intensity: logarithm of the number of abuse-of-power criminal offenders under investigation | 270 | 0.652 | 0.476 | 0 | 1 |
| Scale | Enterprise size: natural logarithm of total firm profit | 270 | 4.162 | 1.403 | 0.060 | 7.421 |
| Expint | Export intensity: ratio of aggregated export value to total firm profit | 270 | 5.038 | 6.788 | 0.101 | 51.402 |
| Rprofit | Profit margin: ratio of total firm profit to total revenue of main businesses | 270 | 16.275 | 9.729 | 4.788 | 89.915 |

## 3. Empirical analysis

### (1) Effects of anticorruption on Innovation Efficiency

First the empirical analysis focuses on the innovation efficiency of anticorruption measures. Ordinary least squares (OLS) regression is conducted based on formula (10). The initial regression results are illustrated in Table 3, which shows that the values of the anticorruption variable (Acor) generally fulfill the expectations and are mostly significant. This shows that, based on the regression equation, anticorruption has significant positive effects on the innovation efficiency of enterprises during both the current and the next periods.

Table 3 Innovation Efficiency of Anticorruption Efforts

| Explanatory variable | Explained variable (Inv) | | | |
|---|---|---|---|---|
| | Current period ($Inv_{i,t}$) | | Next period ($Inv_{i,t+1}$) | |
| Acor | 0.041*** (0.011) | 0.008* (0.004) | 0.0263*** (0.011) | 0.0055 (0.004) |
| Scale | | 0.064*** (0.002) | | 0.061*** (0.002) |
| Expint | | -0.002*** (0.0004) | | -0.0014*** (0.0004) |
| Rprofit | | 0.002*** (0.0002) | | 0.0018*** (0.0002) |
| Const | 0.367*** (0.008) | 0.048*** (0.009) | 0.334*** (0.008) | 0.076*** (0.009) |
| Entity effects | Yes | Yes | Yes | Yes |
| Time effects | No | No | No | No |
| Number of Sample | 270 | 270 | 270 | 270 |
| $R^2$ | 0.0512 | 0.876 | 0.0146 | 0.4941 |

Note: The bracketed values are robust standard errors, whereas ***, **, and * represent significance levels of 1%, 5%, and 10%, respectively.

Normally, at a higher corruption level, anticorruption efforts tend to be more vigorous, in turn exerting greater effects on the innovation activities of enterprises. The ultimate effects on innovation efficiency differ accordingly. As Table 3 does not take into account the effects at different corruption levels, the following section classifies corruption into various levels for a deeper analysis of the effects of anticorruption on innovation efficiency.

**(2) Effects of Anticorruption Effects on Innovation Efficiency at Different Corruption Levels**

To further analyze the effects of anticorruption on innovation efficiency at different corruption levels, corruption above the mean is defined as high-level corruption, and corruption below the mean is defined as low-level corruption. Afterwards, through pairing, the data are divided into two groups-those with high corruption levels and those with low corruption levels. The regression results (as shown in Table 4) are obtained using formula (10) in the model.

Table 4 Innovation Efficiency of Anti-corruption at Different Corruption Levels

| Variable | Innovation efficiency (Inv) | | | |
|---|---|---|---|---|
| | High corruption level | | Low corruption level | |
| | Current period ($Inv_{it}$) | Next period ($Inv_{i,t+1}$) | Current period ($Inv_{it}$) | Next period ($Inv_{i,t+1}$) |
| Acor | ***0.0194***** | ***0.0077*** | ***-0.0145**** | ***-0.0193***** |
| | ***(0.0072)*** | ***(0.0055)*** | ***(0.008)*** | ***(0.0084)*** |
| Scale | 0.0218*** | 0.0299*** | 0.0202*** | 0.0178*** |
| | (0.0037) | (0.0026) | (0.0035) | (0.0052) |
| Expint | -0.0004* | -0.0017*** | -0.0005 | -0.0004 |
| | (0.0004) | (0.0004) | (0.0005) | (0.0008) |
| Rprofit | 0.00055*** | 0.0011*** | 0.0003 | 0.0008 |
| | (0.00019) | (0.0001) | (0.0003) | (0.0006) |
| Const | 0.0245 | 0.0834** | 0.3342*** | 0.3912*** |
| | (0.0523) | (0.0405) | (0.0623) | (0.0597) |
| Time effects | Yes | Yes | Yes | Yes |
| Entity effects | Yes | Yes | Yes | Yes |
| Number of samples | 123 | 84 | 147 | 113 |
| $R^2$ | 0.9819 | 0.9912 | 0.9187 | 0.9310 |

According to the analytical results shown in Table 4, given that attributes, including enterprise size, export intensity, profit margin, time, and geographical region, are controlled for, the coefficients of the anticorruption variable during both the current and the following periods are significantly positive at a high corruption level and significantly negative at a low corruption level. This shows that the anticorruption movement affects enterprises' innovation efficiency positively when there is serious corruption and negatively when corruption is mild. This finding also supports propositions 1 and 2 mentioned in the previous theoretical section. Furthermore, this finding is in line with the findings of a previous study in which low-level corruption was found to be a catalyst for enterprise innovation and high-level corruption was found to be a stumbling block. However, the intensity of the effects is not significant. Therefore, in practice, to improve the innovation efficiency of enterprises, the anticorruption intensity should correspond to the corruption level, and the focus should be placed on diminishing corrupt acts through long-term institutional prevention of corruption.

## 4. Further discussion and robustness testing

The above section provides a preliminary empirical analysis of the effects of anticorruption on enterprises' innovation efficiency considering different corruption levels. This section focuses on robustness analysis for different regions and enterprises.

## (1) Robustness analysis based on different regions

Tables 5 and 6 display the estimated effects of anticorruption on the innovation efficiency of the eastern, central, and western regions at high and low corruption levels, respectively. In general, the results are consistent with the estimations in Table 4, indicating that anticorruption benefits innovation efficiency at a high corruption level and hinders innovation efficiency at a low corruption level. However, the estimated effects are not substantial or statistically significant across different regions. Overall, the estimations are robust and show that the effects of anticorruption on innovation efficiency vary at different corruption levels, but the differences are irrelevant to regional differences. Therefore, in practice, anticorruption intensity should consider differences in corruption levels rather than geographical differences.

Table 5 Robustness analysis based on different regions (at a high corruption level)

| Variable | Innovation efficiency (Inv) | | | | | |
|---|---|---|---|---|---|---|
| | Eastern | | Central | | Western | |
| | Current period | Next period | Current period | Next period | Current period | Next period |
| Acor | ***0.0125*** | ***0.0287*** | ***0.0162***$^{**}$ | ***0.0104*** | ***0.0136*** | ***0.0229*** |
| | ***(0.0192)*** | ***(0.0229)*** | ***(0.008)*** | ***(0.0082)*** | ***(0.0132)*** | ***(0.0113)*** |
| Scale | 0.0102 | 0.0184 | 0.0274$^{***}$ | 0.0255$^{***}$ | 0.009 | 0.0122 |
| | (0.0099) | (0.0118) | (0.0035) | (0.0038) | (0.0121) | (0.0100) |
| Expint | -0.0015 | -0.0022 | 0.0007 | 0.00002 | -0.0007 | -0.0004 |
| | (0.0029) | (0.0029) | (0.0006) | (0.0009) | (0.001) | (0.0008) |
| Rprofit | -0.0004 | 0.0008 | 0.0004$^{**}$ | 0.0005$^{**}$ | 0.00096 | 0.0009 |
| | (0.0027) | (0.0024) | (0.00019) | 0.0002 | (0.00054) | (0.0004) |
| Const | 0.3002$^{*}$ | 0.1512 | -0.0607 | -0.0006 | 0.0223 | -0.0188 |
| | (0.1553) | (0.1981) | (0.0612) | (0.0623) | (0.0865) | (0.0781) |
| Time effects | Yes | Yes | Yes | Yes | Yes | Yes |
| Entity effects | Yes | Yes | Yes | Yes | Yes | Yes |
| Number of samples | 43 | 38 | 55 | 48 | 25 | 22 |
| $R^2$ | 0.9807 | 0.9773 | 0.9941 | 0.993 | 0.9963 | 0.9986 |

Table 6 Robustness analysis based on different regions (at a low corruption level)

| **Variable** | **Innovation efficiency** (Inv) | | | | | |
|---|---|---|---|---|---|---|
| | Eastern | | Central | | Western | |
| | Current period | Next period | Curren t period | Next period | Current period | Next period |
| Acor | ***-0.0169*** | ***-0.017*** | ***-0.0083***$^{**}$ | ***-0.0069***$^{**}$ | ***0.0007*** | ***0.0011*** |
| | ***(0.0165)*** | ***(0.0156)*** | ***(0.0035)*** | ***(0.0029)*** | ***(0.0106)*** | ***(0.0104)*** |
| Scale | 0.0241$^{*}$ | 0.0254$^{**}$ | 0.0039 | 0.0033 | 0.0029 | 0.0001 |
| | (0.0127) | (0.0118) | (0.003) | (0.0026) | (0.0068) | (0.0069) |
| Expint | 0.0015 | 0.0003 | 0.0002 | 0.0001 | 0.0021$^{***}$ | 0.0021$^{***}$ |
| | (0.0025) | (0.0024) | (0.0004) | (0.0004) | (0.0005) | (0.0005) |
| Rprofit | -0.0008 | 0.0002 | 0.00007 | 0.00007 | -0.0002 | -0.0003 |
| | (0.0018) | (0.0017) | (0.0004) | (0.0003) | (0.0002) | (0.0002) |
| Const | 0.4646$^{***}$ | 0.4628$^{***}$ | 0.2209$^{***}$ | 0.2294$^{***}$ | 0.1288 | 0.1403$^{*}$ |
| | (0.1282) | (0.1183) | (0.0279) | (0.0237) | (0.0748) | (0.0729) |
| Time effects | Yes | Yes | Yes | Yes | Yes | Yes |
| Entity effects | Yes | Yes | Yes | Yes | Yes | Yes |
| Number of samples | 77 | 70 | 25 | 24 | 45 | 41 |
| $R^2$ | 0.8736 | 0.8859 | 0.9997 | 0.9997 | 0.9870 | 0.9866 |

## (2) Robustness Analysis Based on Different Enterprise Natures

For a further analysis of the relationship between anticorruption and the innovation efficiency of enterprises of different natures, regression formula (10) is used to examine the state-owned and nonstate-owned sample enterprises at different corruption levels. The results are shown in Table 7. Interestingly, for state-owned enterprises, anticorruption has negative effects on current innovation efficiency at both high and low corruption levels, while it has negative effects on innovation efficiency in the next period at a high corruption level and positive effects at a low corruption level; however, these effects are nonsignificant. The opposite is true for nonstate-owned enterprises, for which anticorruption positively affects innovation efficiency in both the current and the next periods at a high corruption level and negatively affects innovation efficiency at a low corruption level, and the effects are significant. It is found that anticorruption negatively affects the current innovation efficiency of state-owned enterprises, possibly because these enterprises possess closer political connections, and because their innovation investment heavily relies on the government. When anticorruption efforts intensify and place the relationship between government officials and businesses under strict control, the regulation of innovation resources originally obtained through enterprises' political connections also increases; this stongly impacts the innovation of state-owned enterprises, but not nonstate-owned enterprises, which have more independent innovation activities and weaker political connections and rely on an impartial and fair market environment. As the vigorous fight against corruption waged by the Chinese government benefits the maintenance of the impartiality and fairness of the market, it has positive effects on the innovation activities of these nonstate-owned enterprises.

Table 7 Robustness Analysis Based on Different Enterprise Natures

| **Variable** | **Innovation efficiency (Inv)** | | | | | | | |
|---|---|---|---|---|---|---|---|---|
| | State-owned enterprises | | | | Non-state-owned enterprises | | | |
| | High corruption level | | Low corruption level | | High corruption level | | Low corruption level | |
| | Current period | Next period | Current period | Next period | Current period | Next period | Current period | Next period |
| Acor | -0.0043 | -0.0125 | -0.088 | 0.0091 | 0.0175** | 0.0032 | -0.028*** | -0.025** |
| | (0.079) | (0.0084) | (0.061) | (0.0503) | (0.0068) | (0.0062) | (0.0097) | (0.0105) |
| Scale | 0.0497 | 0.0035 | -0.0175 | -0.0024 | 0.0175*** | 0.0369*** | 0.0192*** | 0.0211*** |
| | (0.0321) | (0.0069) | (0.0328) | (0.0309) | (0.0027) | (0.0052) | (0.0051) | (0.0064) |
| Expint | -0.0038 | -0.0004 | -0.0009 | -0.0017 | 0.0003 | -0.0023*** | -0.0003 | -0.0021* |
| | (0.0054) | (0.0011) | (0.0036) | (0.0055) | (0.0005) | (0.0008) | (0.0005) | (0.0011) |
| Rprofit | 0.004 | 0.0001 | -0.00005 | 0.0018 | 0.00097*** | 0.0031*** | 0.0005 | 0.0018** |
| | (0.0033) | (0.0008) | (0.0023) | (0.0041) | (0.0002) | (0.0006) | (0.0003) | (0.0008) |
| Const | -0.0175 | 0.2524*** | 0.9591** | 0.196 | -0.0115 | 0.0123 | 0.4005*** | 0.3822*** |
| | (0.564) | (0.0499) | (0.4416) | (0.3932) | (0.0484) | (0.0369) | (0.0697) | (0.0822) |
| Time effects | Yes | Yes | Yes | Yes | Yes | Yes | Yes | Yes |
| Entity effects | Yes | Yes | Yes | Yes | Yes | Yes | Yes | Yes |
| Number of samples | 106 | 59 | 137 | 99 | 106 | 59 | 137 | 99 |
| $R^2$ | 0.4568 | 0.9866 | 0.2672 | 0.5424 | 0.9860 | 0.9937 | 0.9017 | 0.9271 |

## 5. Conclusion

This study is the first to examine the extent of such effects from multiple perspectives, including overall conditions and different corruption levels, regions, and enterprises. Model estimation and robustness testing reveal that China's anticorruption campaign has positive effects on enterprises' innovation efficiency: Anti-corruption affects innovation efficiency positively at a high corruption level and negatively at a low corruption level. The same is observed regardless of the difference in regions. Nevertheless, the effects differ among enterprises of different natures. For state-owned enterprises, anticorruption has negative impacts on innovation efficiency at both high and low corruption levels, which illustrates that their innovation efficiency relies more heavily on their political connections and, therefore, is under greater impact from the anticorruption movement. For nonstate-owned enterprises, anticorruption affects innovation efficiency positively at a high corruption level and negatively at a low corruption level; this proves that their innovation efficiency is less influenced by the government. Thus, at a high corruption level, the campaign helps ensure fairer treatment for these enterprises in the market and motivates them to improve their innovation efficiency. However, at a low corruption level, increasing the anticorruption effort hinders innovation efficiency for two possible reasons: first, it increases the anticorruption cost allocated to the enterprises and adds to their burden; second, it makes entrepreneurs less inclined to take risk and more prone to pursue stability, causing them to reduce their investment in innovation activities.

In terms of political implications, this study suggests establishing a long-lasting mechanism for corruption prevention, rather than a persistent anticorruption movement because it is evident from the empirical results that the anticorruption movement has negative effects on the innovation efficiency of state-owned enterprises as well as that of nonstate-owned enterprises when there is low-level corruption. In the long run, realizing innovation-driven economic transformation and upgrading requires a long-term preventive mechanism that reduces the occurrence of corrupt acts in advance, instead of a strict punishment mechanism that eliminates corrupt acts after their occurrence. In addition, the government should develop a sound market mechanism, including a modern enterprise system, an intellectual property protection system, and an innovation incentivizing mechanism, to motivate individuals and enterprises to innovate for continuous development and expansion. The empirical results of this study also illustrate that the size of an enterprise has positive effects on its innovation efficiency; in other words, expanding the former improves the latter. Therefore, business expansion can improve innovation efficiency and from a virtuous circle of innovative development. Finally, this study reveals that the effects of anticorruption on innovation efficiency originate from close connections between government officials and businesses, through which the latter can satisfy their own demands and interests via government policies. In other words, political inclination and favoritism exist. A reasonable way to diminish or even root out this issue is to build an impartial, fair, and transparent market mechanism, legal system, and policy use system to safeguard market competition and eliminate any government intervention or protection of competitors, to realize marketplace fairness and free competition.